\documentclass[pageno]{jpaper}

\newcommand{\asplossubmissionnumber}{217}
\pretitle{\begin{center}\normalfont\Large\bfseries}
\posttitle{\end{center}}

\usepackage[normalem]{ulem}

\usepackage{amsmath,amssymb,amsfonts}
\usepackage{color}
\usepackage{epsfig}
\usepackage{makecell}
\usepackage{microtype}
\usepackage{pifont}
\usepackage{xspace}
\usepackage{comment}
\usepackage{listings}
\usepackage{hyperref}

\microtypesetup{
 final,
 tracking=true,
 kerning=true,
 spacing=true,
 factor=1100,
 stretch=20,
 shrink=20
}
\SetTracking{encoding={*}, shape=sc}{0} 

\newcommand{\ignore}[1]{}
\newcommand{\doppio}{\textsc{Doppio}}

\renewcommand{\rothead}[2][60]{\makebox[9mm][c]{\rotatebox{#1}{\makecell[c]{#2}}}}
\newcolumntype{M}{>{\centering\arraybackslash}m{\dimexpr.25\linewidth-2\tabcolsep}}

\newcommand{\cmark}{\ding{51}}

\newcommand{\browsix}{\textsc{Browsix}\xspace}
\newcommand{\Browsix}{\textsc{Browsix}\xspace}

\definecolor{lightgray}{rgb}{.9,.9,.9}
\definecolor{darkgray}{rgb}{.4,.4,.4}
\definecolor{purple}{rgb}{0.65, 0.12, 0.82}

\lstdefinelanguage{JavaScript}{
  keywords={typeof, new, true, false, catch, function, return, null, catch, switch, var, if, in, while
, do, else, case, break},
  keywordstyle=\color{blue}\bfseries,
  ndkeywords={class, export, boolean, throw, implements, import, this},
  ndkeywordstyle=\color{darkgray}\bfseries,
  identifierstyle=\color{black},
  sensitive=false,
  comment=[l]{//},
  morecomment=[s]{/*}{*/},
  commentstyle=\color{purple}\ttfamily,
  stringstyle=\color{red}\ttfamily,
  morestring=[b]',
  morestring=[b]''
}

\begin{document}

\title{\bf \Browsix{}: Bridging the Gap Between Unix and the Browser}

\author{
  Bobby~Powers, John~Vilk, Emery~D.~Berger \\
  University of Massachusetts Amherst
}

\date{}
\maketitle

\thispagestyle{empty}

\begin{abstract}
\vspace{5pt}
Applications written to run on conventional operating systems
typically depend on OS abstractions like processes, pipes, signals,
sockets, and a shared file system. Porting these applications to the
web currently requires extensive rewriting or hosting significant
portions of code server-side because browsers present a nontraditional
runtime environment that lacks OS functionality.

This paper presents \Browsix{}, a framework that bridges the
considerable gap between conventional operating systems and the
browser, enabling unmodified programs expecting a Unix-like
environment to run directly in the browser. \Browsix{} comprises two
core parts: (1) a JavaScript-only system that makes core Unix features
(including pipes, concurrent processes, signals, sockets, and a shared
file system) available to web applications; and (2) extended
JavaScript runtimes for C, C++, Go, and Node.js that support running
programs written in these languages as processes in the
browser. \Browsix{} supports running a POSIX shell,
making it straightforward to connect applications together via pipes.

We illustrate \Browsix{}'s capabilities via case studies that
demonstrate how it eases porting legacy applications to the browser
and enables new functionality. We demonstrate
a \Browsix{}-enabled \LaTeX{} editor that operates by executing
unmodified versions of pdfLaTeX and BibTeX. This browser-only \LaTeX{}
editor can render documents in seconds, making it fast enough to be
practical. We further demonstrate how \Browsix{} lets us port a
client-server application to run entirely in the browser for
disconnected operation. Creating these applications required less than
50 lines of glue code and no code modifications, demonstrating how
easily \Browsix{} can be used to build sophisticated web applications
from existing parts without modification.

\end{abstract}

\section{Introduction}
\label{sec:introduction}

Web browsers make it straightforward to build user interfaces, but
they can be difficult to use as a platform to build sophisticated
applications. Code must generally be written from scratch or heavily
modified; compiling existing code or libraries to JavaScript is not
sufficient because these applications depend on standard OS
abstractions like processes and a shared file system, which browsers
do not support. Many web applications are thus divided between a
browser front-end and a server backend. The server runs on a
traditional operating system, where the application can take advantage
of familiar OS abstractions and run a wide variety of off-the-shelf
libraries and programs.

As a representative example, websites like
ShareLaTeX\footnote{\url{https://www.sharelatex.com/}} and
Overleaf\footnote{\url{https://www.overleaf.com/}} let users write and
edit \LaTeX{} documents in the browser without the need to install a
\TeX{} distribution locally.  This workflow lowers the barrier for
students and first-time \LaTeX{} authors and enables real-time
collaboration, eliminating some of the complexity of creating
multi-author documents. These applications achieve this functionality
by providing a browser-based frontend for editing; user input is sent
to the server for persistence and collaboration purposes. When the
user requests a generated PDF, the website runs \texttt{pdflatex} and
\texttt{bibtex} server-side on the user's behalf, with the resulting
PDF sent to the browser when complete.



These web applications generate PDFs server-side out of necessity
because browsers lack the operating system services and execution
environment that Unix programs expect.  Creating PDFs from \LaTeX{}
requires spawning multiple processes to run \texttt{pdflatex} and
\texttt{bibtex}, which need to read from and write to a shared file
system.  If PDF generation takes too long and the user cancels the
request, the application needs to send a \texttt{SIGTERM} or
\texttt{SIGKILL} signal to clean up any running processes.  If PDF
generation encounters an error, the application needs to pipe the
output of the relevant process back to the client over the network.
Since browsers do not support processes, signals, pipes, sockets, or a
shared filesystem, they cannot perform any of these steps without
program modification.

Previous attempts to cope with this impedance mismatch between
conventional applications and the browser fall short of providing the
environment needed by many programs~(see Section~\ref{sec:related}).
Emscripten and Doppio provide a POSIX-like runtime system for single
processes, including a single-process file system, limited support for
threads, synchronous I/O, and proxying support for TCP/IP
sockets~\cite{doppio:pldi14,emscripten}. While these single-process
runtimes are useful for some applications, they are severely limited
because they are unable to provide the range of operating system
functionality that many legacy applications demand.


To overcome these limitations, we introduce \Browsix{}, a framework
that brings Unix abstractions to the browser through a shared kernel
and common system-call conventions, bridging the gap between
conventional operating systems and the browser. \Browsix{} consists of
two core components: (1) a JavaScript-only operating system that
exposes a wide array of OS services that applications expect
(including pipes, concurrent processes, signals, sockets, and a shared
file system); and (2) extended JavaScript runtimes for C, C++, Go, and
Node.js that let unmodified programs written in these languages and
compiled to JavaScript run directly in the browser. Because
\Browsix{}'s components are written entirely in JavaScript and
require no plugins, applications using \Browsix{} can run in a wide
range of modern web browsers including Google Chrome, Mozilla Firefox, and
Microsoft Edge. \Browsix{} makes it possible to port a wide
range of existing applications and their language runtimes to the
browser by providing the core functionality of a full operating system:

\begin{itemize}

\item \textbf{Processes:} \Browsix{} implements a range of process related
  system calls (including \texttt{fork}, \texttt{spawn}, \texttt{exec}, and
  \texttt{wait4}) and provides a process primitive on top of
  Web Workers, letting applications run in parallel and spawn
  subprocesses.

\item \textbf{Signals:} \Browsix{} supports a substantial subset of the POSIX
  signals API, including \texttt{kill} and signal handlers,
  letting processes communicate with each other
  asynchronously.

\item \textbf{Shared Filesystem:} \Browsix{} lets processes share state through
  a shared FS.

\item \textbf{Pipes:} \Browsix{} exposes the standard \texttt{pipe}
  API, making it simple for developers to compose processes into
  pipelines.

\item \textbf{Sockets:} \Browsix{} supports TCP socket
  servers and clients, making it possible to run server applications
  like databases and HTTP servers together with their clients in the
  browser.

\item \textbf{Language Agnostic:} \Browsix{} includes
  integration with the runtime libraries of Emscripten (C/C++), GopherJS (Go), and
  Node.js (JavaScript) to allow unmodified applications written in these languages
  to run directly as processes in the browser. Through its simple system call API,
  developers can straightforwardly integrate \Browsix{} into additional
  language runtimes.

\end{itemize}
\vspace{10pt}

\begin{figure}[!t]
\centering \includegraphics[width=3.25in]{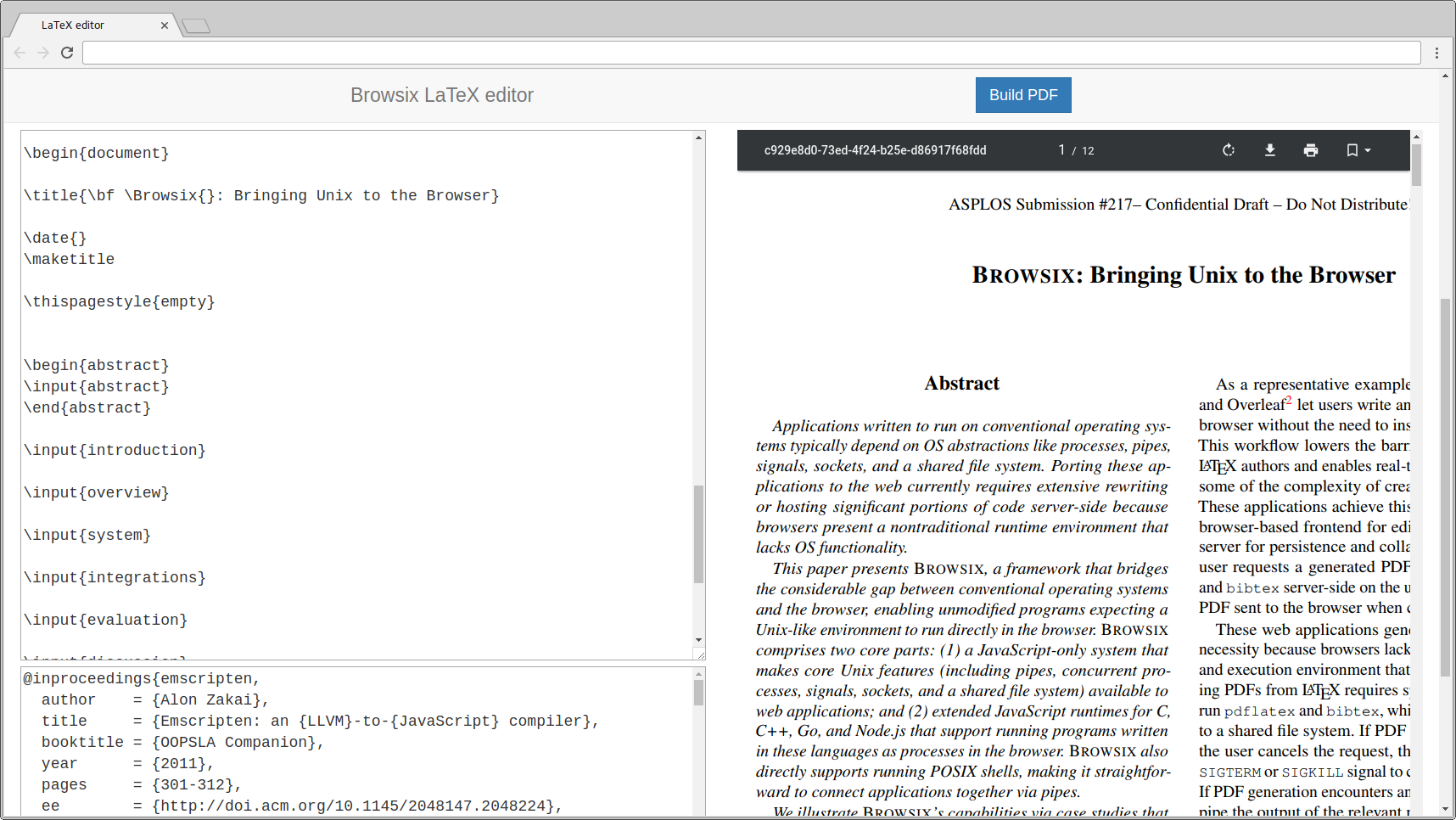}
\caption{A \LaTeX{} editor built using \Browsix{}. \Browsix{}'s OS
  services and language runtimes make it possible to run complex
  legacy code (including \texttt{pdflatex} and \texttt{bibtex})
  directly in the browser, without code modifications (See
  Section~\ref{sec:overview} for details.)}
\label{fig:latexeditor}
\end{figure}

\Browsix{} dramatically simplifies the process of porting complex
applications to the browser environment. As a demonstration, we
have built a \LaTeX{} editor that runs entirely within the browser.
When the user requests a PDF, \Browsix{} runs \texttt{make} to
re-build the document with \texttt{pdflatex} and \texttt{bibtex}, and
pipes their standard output and standard error to the application.
These \TeX{} programs use \Browsix{}'s shared file system to read in
the user's source files, and any packages, class files, and fonts
referenced within, as in a traditional Unix environment.  The
filesystem transparently loads any needed external packages from the
TeX Live distribution over HTTP upon first access.  Subsequent
accesses to the same files are instantaneous, as the browser caches
them.  While a full TeX Live distribution is several gigabytes in
size, a typical paper only needs to retrieve several megabytes worth
of packages before it can be built.  If the user cancels PDF
generation, \Browsix{} sends a \texttt{SIGKILL} signal to these
processes.  If PDF generation fails, the application can display the
captured standard out and standard error.  The result is serverless
PDF generation composed from off-the-shelf parts, with minimal
engineering effort required to glue them together.


We demonstrate the utility of \Browsix{} with two further case
studies. Using \Browsix{}, we build an application that dynamically
routes requests to a remote server or an in-\Browsix{} server, both
compiled from the same source code, depending on the client's
performance and battery characteristics.  We also use \Browsix{} to
build a UNIX terminal exposing a POSIX shell, enabling developers to
launch and compose applications and inspect \Browsix{} state in a
familiar way.

\subsection*{Contributions}

This paper makes the following contributions:

\begin{itemize}

\item \textbf{Bringing OS Abstractions and Services to the Browser.} We demonstrate that it is possible to provide a wide range of
  key Unix abstractions and services in the browser on top of existing
  web APIs. We implement these in \Browsix{}, a JavaScript-only
  framework featuring a kernel and system calls that runs on all
  modern browsers (\S\ref{sec:browsix}).

\item \textbf{Runtime Integration for Existing Languages.} We extend the JavaScript runtimes of Emscripten (a C/C++ to
  JavaScript toolchain), GopherJS (a Go to JavaScript compiler), and
  Node.js with \Browsix{} support, letting unmodified C, C++, Go, and
  Node.js programs execute and interoperate with one another within
  the browser as \Browsix{} processes (\S\ref{sec:integrations}).

\item \textbf{Case Studies.} We demonstrate \Browsix{}'s utility by building a \LaTeX{}
  editor, a serverless client-server web application, and a Unix
  terminal out of off-the-shelf components without modification.  We
  characterize \Browsix{}'s performance under these case studies and
  with microbenchmarks~(\S\ref{sec:eval}) and show that its overhead
  is low enough for real-world usage.

\item \textbf{Guidance for Future Browsers.} Based on our experience writing \Browsix{}, we discuss current
  browser limitations and propose solutions~(\S\ref{sec:discussion}).

\end{itemize}

\section{\Browsix{} Overview}
\label{sec:overview}

To give an overview of \Browsix{}'s features, this section walks through the
process of using \Browsix{} to build an in-browser \LaTeX{} editor
using TeX Live utilities and GNU Make.
Figure~\ref{fig:latexeditor} displays a screenshot of the editor.

\subsection{\LaTeX{} Editor Overview}

The editor presents a split-screen view to the user, with the
document's \LaTeX{} source on the left, and generated PDF preview on
the right.  The editor's UI is a standard web application, and
represents the only new code.  When the user clicks on the ``Build
PDF'' button, the editor uses \Browsix{} to invoke GNU Make in a
\Browsix{} process, which rebuilds the PDF.

The process for building the PDF is familiar to anyone who has used
\LaTeX{}, except \Browsix{} performs all of the needed steps entirely
in the browser instead of server-side. It runs GNU Make to read a
Makefile from \Browsix{}'s file system, which contains rules for
rebuilding \LaTeX{} projects. Make then runs \texttt{pdflatex} and
\texttt{bibtex}, depending on whether the user has updated the
references file.

\texttt{pdflatex} and \texttt{bibtex} read any required \LaTeX{}
packages, fonts, and other system files from \Browsix{}'s file system,
which lazily pulls in files as needed from the network. Both of these
applications write their output to \Browsix{}'s file system.

Once all steps have completed (or an error has occurred), the Make
process exits with an exit code indicating whether or not PDF
generation succeeded.  \Browsix{} sends the exit code back to the web
application.  If GNU Make exits normally, the editor reads the PDF
from \Browsix{}'s shared filesystem and displays it to the user.
Otherwise, it displays the standard output from \texttt{pdflatex} and
\texttt{bibtex} to the user, which describes the source of the error.

\subsection{Building with \Browsix{}}

Building any web application that runs Unix programs in \Browsix{}
generally consists of the same three step process: (1) compile the
programs to JavaScript (using tools with \Browsix support), (2) stage files
required by the application for placement in the in-browser
filesystem, and (3) add setup code to the web application to initiate
\Browsix{} and launch the programs.


\paragraph{Compiling to JavaScript:}

To run \texttt{pdflatex}, \texttt{bibtex}, and GNU Make in \Browsix{},
the developer compiles each program to JavaScript using Emscripten, a
C/C++ to JavaScript compiler~\cite{emscripten}.  We extend
Emscripten's runtime library with \Browsix support, so standard Unix
APIs map to \Browsix primitives.  We discuss this extension in more
detail in Section~\ref{subsec:emscripten}.

Before compilation, the developer needs to determine if any of the
programs use the \texttt{fork} command.  Due to browser limitations
explored in Section~\ref{par:stack-management}, \Browsix{} can only implement
\texttt{fork} as an asynchronous system call, which requires the use
of a variant of Emscripten's compilation procedure that generates less
efficient code.  If this option is configured incorrectly, the program
will fail at runtime when it attempts to invoke \texttt{fork}. For the
\LaTeX{} example, only GNU Make uses \texttt{fork} and requires
this setting.

From this point, the build process is mostly unchanged from a standard
compilation.  For programs that use the autotools build system, such
as GNU Make and TeX Live, instead of running \texttt{./configure}, the
developer invokes \texttt{emconfigure ./configure}.  This
wrapper overrides standard tools like \texttt{cc} and \texttt{ar} with
\texttt{em} prefixed alternatives that compile the program with
Emscripten, which produces individual JavaScript files for each
program.

\paragraph{Staging the Filesystem:}
Next, the developer configures \Browsix{}'s in-browser filesystem so
that it hosts all of the files that the programs require.
\Browsix{}'s file system extends Doppio's BrowserFS file system with
multi-process support, building on its support for files backed by cloud
storage, browser-local storage, traditional HTTP servers, and
more~\cite{doppio:pldi14}.

For our \LaTeX{} example, both \texttt{pdflatex} and \texttt{bibtex}
require read access to class, font, and other files from a \LaTeX{}
distribution to function properly. While a complete TeX Live distribution
contains over 60,000 individual files, the average \LaTeX{}
document only references a small subset of these files.

To reduce load times and minimize the amount of storage required
on a client's device, the developer can leverage \Browsix{}'s
filesystem to load only the needed files. In this case, the developer
uploads a full TeX Live distribution to an HTTP server and configures
\Browsix{}'s filesystem to use an HTTP-backed filesystem backend.  The
filesystem will then load these files on-demand from the network upon
first access.  The browser caches these files automatically, making
subsequent access much faster.

\paragraph{\Browsix{} Setup Code:}

Finally, the developer adds code to the web application to load and
initialize \Browsix{}, and to launch \texttt{make} to build the PDF.
A \texttt{script} tag in the HTML loads \texttt{browsix.js}, and a
subsequent \texttt{script} tag with inline JavaScript calls
\browsix{}'s \texttt{Boot} function with the desired filesystem
configuration.

Additional application-specific initialization follows as
usual. Once the filesystem is ready, the developer adds code to
read the contents of \texttt{main.tex} and \texttt{main.bib} from
\Browsix{}'s filesystem, and display the contents in the editor;
The application then registers a callback function with the ``Build PDF''
button to be run whenever the user clicks the button.

\subsection{Execution with \Browsix{}}

When the application's callback is executed in response to a user's
``Build PDF'' click, the application invokes the \texttt{system} method on
its kernel instance to start \texttt{make}.  Make runs \texttt{pdflatex} and
\texttt{bibtex} as described in Section~\ref{sec:introduction}.  When
the application receives a notification from \Browsix{} that Make has
exited, it inspects Make's exit code.  If it is zero, the PDF was
generated successfully and is read from the filesystem.  Otherwise,
the captured standard output and standard error are displayed to the
user so they can debug their markup.

This overview demonstrates how straightforward \Browsix{} makes it to
run existing components -- designed to work in a Unix environment --
and execute them seamlessly inside a web browser. The next two sections
provide technical details on how \Browsix{} provides Unix-like
abstractions in the browser environment and integrates with language runtimes.

\begin{figure}[!t]
\centering
\begin{tabular}{lrr}
\textbf{Component}     & \textbf{Lines of Code (LoC)} \\
\hline
{Kernel}                  & 2,249  \\
{BrowserFS modifications} & 1,231  \\
{Shared syscall module}   & 421    \\
{Emscripten integration*} & 1,557  \\ 
\textit{(C/C++ support)}  &        \\
{GopherJS integration*}   & 926    \\ 
\textit{(Go support)}     &        \\
{Node.js integration*}    & 1,742  \\ 
\hline
{\textbf{\small{TOTAL}}}  & \textbf{8,126} \\
\end{tabular}
\caption{\Browsix{} components.  * indicates these
	components are written in JavaScript, while the rest of the
	components are written in TypeScript (which compiles to JavaScript).}
\label{fig:linecount}
\end{figure}

\section{\Browsix{} OS Support}
\label{sec:browsix}

The core of \Browsix{}'s OS support is a kernel that controls access
to shared Unix services. Unix services, including the shared file system,
pipes, sockets, and task structures, live inside the kernel, which
runs in the main browser thread.  Processes run separately and in
parallel inside Web Workers, and access \Browsix{} kernel services
through a system call interface.  \Browsix{} and all of its runtime
services are implemented in JavaScript and TypeScript, a typed variant
of JavaScript that compiles to pure
JavaScript. Figure~\ref{fig:linecount} provides a breakdown of each of
\Browsix{}'s components.

\subsection{Kernel}

The kernel lives in the main JavaScript context alongside the web
application, and acts as the intermediary between processes and
loosely coupled Unix subsystems.  Processes issue system calls to
the kernel to access shared resources, and the kernel relays these
requests to the appropriate subsystem.  When the subsystem responds to
the system call, it relays the response to the process.  The kernel is
also responsible for dispatching signals to processes, which we
describe further in Section~\ref{sec:processes}.
Figure~\ref{system:calls} presents a partial list of the system
calls that the kernel currently supports.


In a departure from modern Unix systems, \Browsix{} does not support
multiple users. A traditional kernel would, for example, use user
identities to check permissions on certain system calls or for access
to files. Instead, \Browsix{} leverages and relies on the browser's
built-in sandbox and security features, such as the same origin
policy. In other words, a \Browsix{} application enjoys the same level
of protection and security as any other web application.

\subsection{System Calls}
\label{sec:syscalls}

\begin{figure}[!t]
\centering
\begin{tabular}{llr}
\textbf{Class}      & \textbf{System calls} \\
\hline
{Process Management} & \small \texttt{fork}, \texttt{spawn}, \texttt{pipe2}, \\
                     & \small \texttt{wait4}, \texttt{exit} \\[1ex]
{Process Metadata}   & \small \texttt{chdir}, \texttt{getcwd}, \texttt{getpid} \\[1ex]
{Sockets}            & \small \texttt{socket}, \texttt{bind}, \texttt{getsockname}, \\
                     & \small \texttt{listen}, \texttt{accept}, \texttt{connect}    \\[1ex]
{Directory IO}       & \small \texttt{readdir}, \texttt{getdents}, \\
                     & \small \texttt{rmdir}, \texttt{mkdir} \\[1ex]
{File IO}            & \small \texttt{open}, \texttt{close}, \texttt{unlink}, \\
                     & \small \texttt{llseek}, \texttt{pread}, \texttt{pwrite} \\[1ex]
{File Metadata}      & \small \texttt{access}, \texttt{fstat}, \texttt{lstat}, \\
                     & \small \texttt{stat}, \texttt{readlink}, \texttt{utimes}
\end{tabular}
\caption{A representative list of the system calls implemented by the
  \Browsix{} kernel. \texttt{fork} is only supported for C and C++
  programs.}
\label{system:calls}
\end{figure}

The \Browsix{} kernel supports two types of system calls: asynchronous
and synchronous.  Asynchronous system calls work in all modern
browsers, but impose a high performance penalty on C and C++ programs.
Synchronous system calls enable C and C++ programs to perform much
better, but currently only work in Google Chrome via a mechanism we
describe below; this mechanism is already standardized and is on track to be
supported by other mainstream browsers.

\paragraph{Asynchronous System Calls:}
\Browsix{} implements asynchronous system calls in a
continuation-passing style (CPS). A process initiates a system call by
sending a message to the kernel with a process-specific unique ID, the
system call number, and arguments. \Browsix{} copies all arguments,
such as file descriptor or a buffer to write to a file
descriptor, from the process to the kernel - no memory is shared.
When the kernel sends a response, the Web Worker process executes the
continuation (or callback) with response values, also copied from the
kernel's heap into the process's heap.

Asynchronous system calls work well for Node.js and Go, but are a poor
match for many C and C++ programs.  In Node.js, filesystem and other
APIs accept a callback parameter to invoke when a response is ready,
which matches \Browsix{}'s asynchronous system call mechanism.  The
GopherJS runtime provides support for suspending and resuming the call
stack in order to implement goroutines (lightweight thread-like
primitives); this support also meshes with \Browsix{}'s
approach. However, when using Emscripten to compile C and C++
programs, they must be compiled in an interpreted mode (called the
Emterpreter) in order to use asynchronous system calls. This mode
produces much less efficient code than the standard compiler, which
produces \texttt{asm.js} output by default.

\paragraph{Synchronous System Calls:}
Synchronous system calls work by sharing a view of a process's address
space between the kernel and the process, similar to a traditional
operating system kernel like Linux.  At startup, the language runtime
in a process wishing to use synchronous system calls passes to the kernel
(via an asynchronous system call) a
reference to the heap (a SharedArrayBuffer object), along with two offsets into the heap:
where to put system call return values, and an offset to use to wake
the process when the syscall is complete.

A process invokes a synchronous system call by sending a message, as
in the asynchronous case, but arguments are just integers and integer
offsets (representing pointers) into the shared memory array, rather
than larger objects (like ArrayBuffers) that would need to be copied
between heaps.  For system calls like \texttt{pread}, data is
copied directly from the filesystem, pipe or socket into the process's
heap, avoiding a potentially large allocation and extra copy.

After sending a message to the kernel, the process performs a blocking
wait on the address previously arranged with the kernel and is awakened
when the system call has completed or a signal is received.  This wait
is provided by the JavaScript \texttt{Atomics.wait} function, part of
the ECMAScript Shared Memory and Atomics
specification~\cite{tc39:2016sab}.  A side effect of this approach is
that \texttt{fork} is not compatible with synchronous system calls, as
there is no way to re-wind or jump to a particular call stack in the
child Web Worker.

Synchronous system calls are faster in practice for a number of
reasons. First, they only require one message to be passed between the
kernel and process, which is a relatively slow operation.  Second,
system call arguments are numbers, rather than potentially large
arrays that need to be copied between JavaScript contexts.  Finally,
synchronous system calls provide a blocking primitive and do not
depend on language runtimes to unwind and rewind the call stack.  As
such, they are suitable for use with \texttt{asm.js} and WebAssembly
functions on the call stack, which are faster and more amenable to
optimization by the JavaScript runtime than Emscripten's interpreter.

Synchronous system calls currently require the in-development browser
features SharedArrayBuffers and Atomics, and currently only work in
Google Chrome when launched with extra flags.  SharedArrayBuffers and
Atomics are on the standards track, and are expected to be supported
un-flagged in mainstream browsers in the near future.

\subsection{Processes}
\label{sec:processes}

\Browsix{} relies on \emph{Web Workers} as the foundation for emulating Unix
processes. However, Web Workers differ substantially from Unix processes,
and \Browsix{} must provide a significant amount of functionality to
bridge this gap.

In Unix, processes execute in isolated virtual address spaces, run in
parallel with one another when the system has multiple CPU cores, and
can interact with system resources and other processes via system
calls. However, the web browser does not expose a process API to web
applications.  Instead, web applications can spawn a Web Worker
that runs a JavaScript file in parallel with the application.

A Web Worker has access to only a subset of browser interfaces
(notably excluding the Document Object Model (DOM)), runs in a
separate execution context, and can only communicate with the main
browser context via asynchronous message passing. Web Workers are not
aware of one another, cannot share memory with one another, and can
only exchange messages with the main browser context that created them
(see Section~\ref{sec:system:limitations} for a discussion). Major browsers like Chrome and Safari
do not support spawning sub-workers from workers, so-called
\emph{nested workers}, and have not added support for them since they
were first proposed in 2009.  Thus, if a Web Worker needs to perform a
task in parallel, it must delegate the request to the main browser
thread, and proxy all messages to that worker through the main browser
thread. Perhaps unsurprisingly, the limitations and complexity of Web
Workers have hindered their adoption in web applications.

By contrast, \Browsix{} implements Unix processes on top of Web
Workers, giving developers a familiar and full-featured abstraction
for parallel processing in the browser.  Each \Browsix{} process has
an associated task structure that lives in the kernel that contains
its process ID, parent's process ID, Web Worker object, current
working directory, and map of open file descriptors.  Processes have
access to the system calls in Figure~\ref{system:calls}, and invoke
them by sending a message with the system call name and arguments to
the kernel.  As a result, processes can share state via the file
system, send signals to one another, spawn sub-processes to perform
tasks in parallel, and connect processes together using pipes.  Below,
we describe how \Browsix{} maps familiar OS interfaces onto Web
Workers.

\paragraph{\texttt{spawn}:}
\Browsix{} supports \texttt{spawn}, which constructs a new process
from a specified executable on the file system.  \texttt{spawn} is the
primary process creation primitive exposed in modern programming
environments such as Go and Node.js, as \texttt{fork} is unsuitable
for general use in multithreaded processes. \texttt{spawn} lets a
process specify an executable to run, the arguments to pass to that
executable, the new process's working directory, and the resources
that the subprocess should inherit (such as file descriptors).  In
\Browsix{}, executables include JavaScript files, file beginning with
a shebang line, and WebAssembly files.  When a process invokes
\texttt{spawn}, \Browsix{} creates a new task structure with the
specified resources and working directory, and creates a new Web
Worker that runs the target executable or interpreter.

There are two technical challenges to implementing
\texttt{spawn}. First, the Web Worker constructor takes a URL to a
JavaScript file as its first argument.  Files in \Browsix{}'s file
system may not correspond to files on a web server. For example, they
might be dynamically produced by other \Browsix{} processes. To work
around this restriction, \Browsix{} generates a JavaScript \emph{Blob}
object that contains the data in the file, obtains a
dynamically-created \emph{URL} for the blob from the browser's window
object, and passes that URL as a parameter to the Web Worker
constructor. All modern web browsers now support constructing Workers
from blob URLs.

The second challenge is that there is no way to pass data to a Worker
on startup apart from sending a message.  As processes synchronously
access state like the arguments vector and environment map,
\Browsix{}-enabled runtimes delay execution of a
process's \texttt{main()} function until after the worker has received
an ``init'' message containing the process's arguments and
environment.

\paragraph{\texttt{fork}:}
The \texttt{fork} system call creates a new process containing a copy
of the current address space and call stack.  Fork returns twice --
first with a value of zero in the new process, and with the PID of the
new process in the original.  Web Workers do not expose a cloning
API, and JavaScript lacks the reflection primitives required to
serialize a context's entire state into a snapshot.  Thus, \Browsix{}
only supports \texttt{fork} when a language runtime is able to completely
enumerate and serialize its own state. Section~\ref{sec:integrations}
describes how we extend Emscripten to provide \texttt{fork} support for
C/C++ programs compiled to JavaScript.

\paragraph{\texttt{wait4}:}
The \texttt{wait4} system call reaps child processes that have finished executing.
It returns immediately if the specified child has already exited, or the
\texttt{WNOHANG} option is specified.  Waiting requires that the
kernel not immediately free task structures, and required us to
implement the zombie task state for children that have not yet been
waited upon.  The C library used by Emscripten, musl, uses the
\texttt{wait4} system call to implement the C library functions
\texttt{wait}, \texttt{wait3}, and \texttt{waitpid}.

\paragraph{\texttt{exit}:}
Language runtimes with \Browsix-support are required to explicitly issue an
\texttt{exit} system call when they are done executing, as the containing
Web Worker context has no way to know that the process has finished.
This is due to the event-based nature of JavaScript environments --
even if there are no pending events in the Worker's queue, the main
JavaScript context could, from the perspective of the browser, send
the Worker a message at any time.

\paragraph{\texttt{getpid}, \texttt{getppid}, \texttt{getcwd}, \texttt{chdir}:}
These four system calls operate on the data in current process's task
structure, which lives in the \Browsix{} kernel. \texttt{getpid} returns
the process's ID, \texttt{getppid} returns the parent process's ID,
\texttt{getcwd} returns the process's working directory, and
\texttt{chdir} changes the process's working directory.

\subsection{Pipes}

\Browsix{} pipes are implemented as in-memory buffers with read-side
wait queues.  If there is no data to be read when a process issues a
\texttt{read} system call, \Browsix{} enqueues the callback
encapsulating the system call response which it invokes when data is
written to the pipe.  Similarly, if there is not enough space in a
pipe's internal buffer, \Browsix{} only invokes the callback
encapsulating the system call response to the write operation when the
pipe is read from.

\subsection{Sockets}

\Browsix{} implements a subset of the BSD/POSIX socket API, with
support for \texttt{SOCK\_STREAM} (TCP) sockets for communicating
between \Browsix{} processes.  These sockets enable servers that
\texttt{bind}, \texttt{listen} and then \texttt{accept} new
connections on a socket, along with clients that \texttt{connect} to a
socket server, with both client and server reading and writing from
the connected file descriptor.  Sockets are sequenced, reliable,
bi-directional streams.

\subsection{Shared File System}

\Browsix{} builds on and significantly extends BrowserFS's file
system, part of Doppio~\cite{doppio:pldi14}. BrowserFS already included
support for multiple mounted filesystems in a single hierarchical
directory structure.  BrowserFS provides multiple file system backend
implementations, such as in-memory, zip file, XMLHttpRequest, Dropbox,
and an overlay filesystem.  BrowserFS provides a unified, encapsulated
interface to all of these backends.

\Browsix{} extends BrowserFS in two key ways: it adds multi-process
support and incorporates improved support for loading files over
HTTP. To provide multi-process support, \Browsix{}'s file system adds
locking operations to the overlay filesystem to prevent operations
from different processes from interleaving.  In addition, \Browsix
incorporates domain-specific optimizations into its file system; for
example, it avoids expensive operations like recording the call stack
when a path lookup fails (a common event).

\Browsix{} modifies BrowserFS's overlay backend to lazily load files
from its read-only underlay; the original version eagerly read all
files from the read-only filesystem upon initialization. \Browsix{}'s approach
drastically improves the startup time of the kernel, minimizes the
amount of data transferred over the network, and enables applications
like the \LaTeX{} editor where only a small subset of files are
required for a given end user.

Finally, \Browsix{} implements system calls that operate on paths, like
\texttt{open} and \texttt{stat}, as method calls to the kernel's
BrowserFS instance.  When a system call takes a file descriptor as an
argument, the kernel looks up the descriptor in the tasks's file
hashmap and invokes the appropriate method on that file
object, calling into BrowserFS for regular files and directories. Child processes inherit file descriptor tables, and \Browsix{}
manages each object (whether it is a file, directory, pipe or socket)
with reference counting.

\section{\Browsix{} Runtime Support}
\label{sec:integrations}


Applications access \Browsix{} system calls indirectly through their
runtime systems.  This section describes the runtime support we added
to GopherJS, Emscripten, and Node.js along with the APIs exposed to
web applications so they can execute programs in \Browsix{}.


\subsection{Browser Environment Extensions}
\label{subsec:browser}

Web applications run alongside the \Browsix{} kernel in the main browser context, and have access to \Browsix{}
features through several global APIs.
\Browsix{} exposes new APIs for process creation,
file access, and socket notifications, and an XMLHttpRequest-like interface to send HTTP requests to \Browsix{} processes.

File access acts as expected, and allows the client to manipulate the
filesystem, invoke a utility or pipeline of utilities, and read state
from the filesystem after programs have finished executing.
Figure~\ref{figure:createproc} shows how client applications invoke
\Browsix{} processes and react when processes exit through an API
similar to C's \texttt{system}.

\begin{figure}
\begin{lstlisting}[language=JavaScript]
kernel.system(
    'pdflatex example.tex',
    function(pid, code) {
        if (code === 0) {
            displayPDF();
        } else {
            displayLatexLog();
        }
    }, logStdout, logStderr);
\end{lstlisting}
\caption{Creating a \Browsix{} process from JavaScript.}
\label{figure:createproc}
\end{figure}

Socket notifications let applications register a callback to be
invoked when a process has started listening on a particular port.
These notifications let web applications launch a server as a process and appropriately delay
communicating with the server until it is listening for messages.
Web applications do not need to resort to polling or ad hoc waiting.

\Browsix provides an XMLHttpRequest-like API for sending requests from
the web application to in-browser HTTP servers running in \Browsix{}.
This allows JavaScript to interact with HTTP 1.1 servers running as \Browsix{}
processes as if they were remote HTTP servers.  The API encapsulates
the details of connecting a \Browsix{} socket to the server,
serializing the HTTP request to a byte array, sending the byte array
to the \Browsix{} process, processing the (potentially chunked)
HTTP response, and generating the expected web events.

\subsection{Common Services}
\label{subsec:syscall}

\Browsix{} provides a small \texttt{syscall} layer as a JavaScript
module that runs in a Web Worker. This layer provides a concrete,
typed API for asynchronous system calls over the browser's message
passing primitives.  Language runtimes use this module to communicate
with the shared kernel.  Methods provided by the \texttt{syscall}
layer take the same arguments as Linux system calls of the same name,
along with an additional argument: a callback function.  This callback
is executed when the \texttt{syscall} module receives a message
response from the kernel.  Unlike a traditional single-threaded
process, a \Browsix{} process can have multiple outstanding system
calls, which enables runtimes like GopherJS to implement user-space
threads on top of a single Web Worker execution context.

Signals are sent over the same message passing interface as system
calls.  The common \texttt{syscall} module provides a way to register
signal handlers for the standard Unix signals, such as
\texttt{SIGCHLD}.

\subsection{Runtime-specific Integration}

For many programming languages, existing language runtimes targeted
for the browser must bridge the impedance mismatch between synchronous
APIs present on Unix-like systems and the asynchronous
world of the browser. Compile-to-JavaScript
systems like Emscripten, ClojureScript~\cite{clojurescript},
Scala.js~\cite{scala-js}, js\_of\_ocaml~\cite{js-of-ocaml}, WhaleSong (Racket)~\cite{whalesong},
and GopherJS all employ different approaches. Since \Browsix{} supports
both synchronous and asynchronous system calls, language runtimes
can choose the system call convention most appropriate for their
implementation.

This section describes the runtime support we added to language runtimes for
Go, C/C++, and Node.js. Extending \Browsix{} support
to additional language runtimes remains as future work.

\paragraph{Go:}
\label{subsec:go}
Go is a systems language developed at Google designed for readability,
concurrency, and efficiency. To run Go programs under \Browsix{}, we
extended the existing GopherJS compiler and runtime to support issuing
and waiting for system calls under \Browsix{}.  GopherJS already
provides full support for Go language features like goroutines
(lightweight threads), channels (communication primitives), and
delayed functions.

\begin{figure}
\begin{lstlisting}[language=JavaScript]
function sys_getdents64(cb, trap, fd, dirp, len) {
    var done = function (err, buf) {
        if (!err)
            dirp.set(buf);
        cb([err ? -1 : buf.byteLength, 0, err ? err : 0]);
    };
    syscall_1.syscall.getdents(fd, len, done);
}
\end{lstlisting}
\caption{Implementing the \texttt{getdents64} system call in GopherJS.}
\label{figure:integration-go}
\end{figure}

We extended the GopherJS runtime with support for \Browsix{} through
modifications to the runtime.  The main integration points are a
\Browsix{}-specific implementation of the \texttt{syscall.RawSyscall}
function (which handles syscalls in Go), along with overrides of
several Go runtime functions.

The replacement for \texttt{RawSyscall} is implemented in Go. It allocates a
Go channel object, and this function invokes the \Browsix{} JavaScript
\texttt{syscall} library, passing the system call number, arguments, and a
callback to invoke.  \texttt{RawSyscall} then performs a blocking read
on the Go channel, which suspends the current goroutine
until the callback is invoked.  When the system call response is
received from the \Browsix{} kernel, GopherJS's existing runtime takes
care of re-winding the stack and continuing execution.  The \texttt{syscall}
library invokes a function specific to each supported system call to marshal data to
and from the \Browsix{} kernel.  Adding support for any new system
call is a matter of writing a small handler function and registering
it; an example is shown in
Figure~\ref{figure:integration-go}

\Browsix{} replaces a number of low-level runtime functions; the most
important are \texttt{syscall.forkAndExecInChild} and
\texttt{net.Listen}.  The former is overridden to directly invoke
\Browsix's \texttt{spawn} system call, and the latter to provide
access to \Browsix socket services.  Additional integration points
include an explicit call to the \texttt{exit} system call when the main function
exits, and waiting until the process's arguments and environment have
been received before starting \texttt{main()} (see
\S\ref{sec:processes}).

\begin{figure}
\begin{lstlisting}[language=JavaScript]
  __syscall220: function(which, varargs) {
#if EMTERPRETIFY_ASYNC
      return EmterpreterAsync.handle(function(resume) {
        var fd = SYSCALLS.get(), dirp = SYSCALLS.get(), count = SYSCALLS.get();
        var done = function(err, buf) {
          if (err > 0)
            HEAPU8.subarray(dirp, dirp+buf.byteLength).set(buf);
          resume(function() {
            return err;
          });
        };
        SYSCALLS.browsix.syscall.async(done, 'getdents', [fd, count]);
      });
#else
      var fd = SYSCALLS.get(), dirp = SYSCALLS.get(), count = SYSCALLS.get();
      return SYSCALLS.browsix.syscall.sync(220, fd, dirp, count);
#endif
  },
\end{lstlisting}
\caption{Implementing the \Browsix{} \texttt{getdents64} syscall in Emscripten.}
\label{figure:integration-c}
\end{figure}

\paragraph{C and C++:}
\label{subsec:emscripten}
We also extend Emscripten, Mozilla Research's LLVM-based C and C++
compiler that targets JavaScript, with support for \Browsix{}.
\Browsix-enhanced Emscripten supports two modes - synchronous system
calls and asynchronous system calls (described in
Section~\ref{sec:syscalls}), one of which is selected at compile time.
When asynchronous system calls are used, it requires use of
Emscripten's interpreter mode (named the ``Emterpreter'') to save
and restore the C stack.  \Browsix{}'s asynchronous system calls require
all functions that may be on the stack under a system call to be
interpreted so that the stack can be restored when the system call
completes. Emscripten can selectively compile other parts of an
application to \texttt{asm.js}, which will be JIT-compiled and run as
native JavaScript by the browser.  Synchronous system calls do not
have this limitation.

As with GopherJS, Emscripten provides a clear integration point at the
level of system calls.  Emscripten provides implementations for a
number of system calls, but is restricted to performing in-memory
operations that do not block.  We replace Emscripten system call
implementations with ones that call into the \Browsix{} kernel, such
as in Figure~\ref{figure:integration-c}.  In the case of
\texttt{getdents} and \texttt{stat}, padding was added to C structure
definitions to match the layout expected by the \Browsix{} kernel.

When a C process invokes \texttt{fork}, the runtime sends a copy of the
global memory array, which includes the C stack and heap, along with
the current program counter (PC) to the kernel.  After the kernel
launches a new Web Worker, it transfers this copy of global memory and
PC to the new Worker as part of the initialization message.  When
the Emscripten runtime in the new \Browsix process receives the
initialization message, if a memory array and PC are present the
runtime swaps them in and invokes the Emterpreter to continue from
where \texttt{fork} was invoked.

\paragraph{Node.js:}
\label{subsec:node}
Node.js (a.k.a. ``Node'') is a platform for building servers and command line
tools with JavaScript, implemented in C, C++ and JavaScript, on top of
the v8 JavaScript engine.  Node.js APIs are JavaScript modules that
can be loaded into the current browser context by invoking the
\texttt{require} built-in function.  These high-level APIs are
implemented in platform-agnostic JavaScript, and call into lower-level
C++ bindings, which in turn invoke operating system interfaces like
filesystem IO, TCP sockets, and child process management.  Node.js
embraces the asynchronous, callback-oriented nature of JavaScript --
most Node APIs that invoke system calls take a callback parameter that
is invoked when results are ready.

To run servers and utilities written for Node.js under \Browsix{}, we
provide a \texttt{browser-node} executable that packages Node's
high-level APIs with pure-JavaScript replacements for Node's C++
bindings that invoke \Browsix{} system calls as a single file that
runs in a \Browsix{} process. \Browsix{} also replaces several other
native modules, like the module for parsing and generating HTTP
responses and requests, with pure JavaScript implementations. Node
executables can be invoked directly, such as \texttt{node server.js},
or will be invoked indirectly by the kernel if node is specified as
the interpreter in the shebang line of a text file marked as
executable.

\section{Evaluation}
\label{sec:eval}

Our evaluation answers the following questions: (1) Does bringing
Unix abstractions into the browser enable compelling use cases?  (2)
Is the performance impact of running programs under \Browsix
acceptable?

\subsection{Case Studies}

We evaluate the applicability and advantages of bringing Unix
abstractions into the browser with two case studies, in addition to
the \LaTeX{} editor from the overview (\S\ref{sec:overview}).  First,
we build a web application for creating memes that can run its unmodified
server in \browsix{}. The meme generator transparently switches between generating memes in-browser or
server-side depending on network and device characteristics.
Second, we build a Unix terminal that lets application developers use \texttt{dash}, a widely-used POSIX shell,
to interact with \Browsix{} in a familiar manner.

\begin{figure}[!t]
\centering
\includegraphics[width=0.48\textwidth]{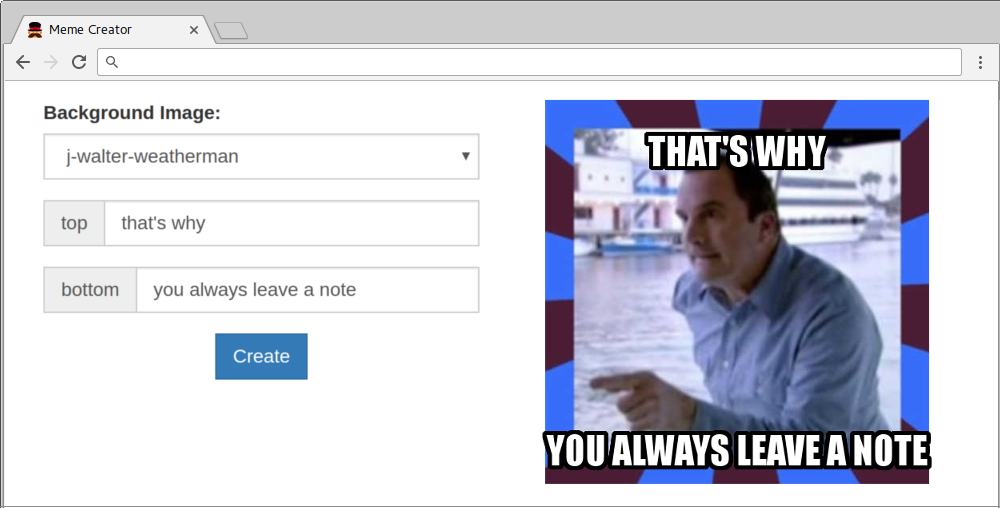}
\caption{A meme generator built using \Browsix{}. All server-side functionality was moved into the browser without modifying any code.}
\label{fig:imagemacrogen}
\end{figure}

\begin{figure*}[!t]
  \centering\subfloat[Meme creator running without \Browsix{}]{
    \includegraphics[width=.7\textwidth]{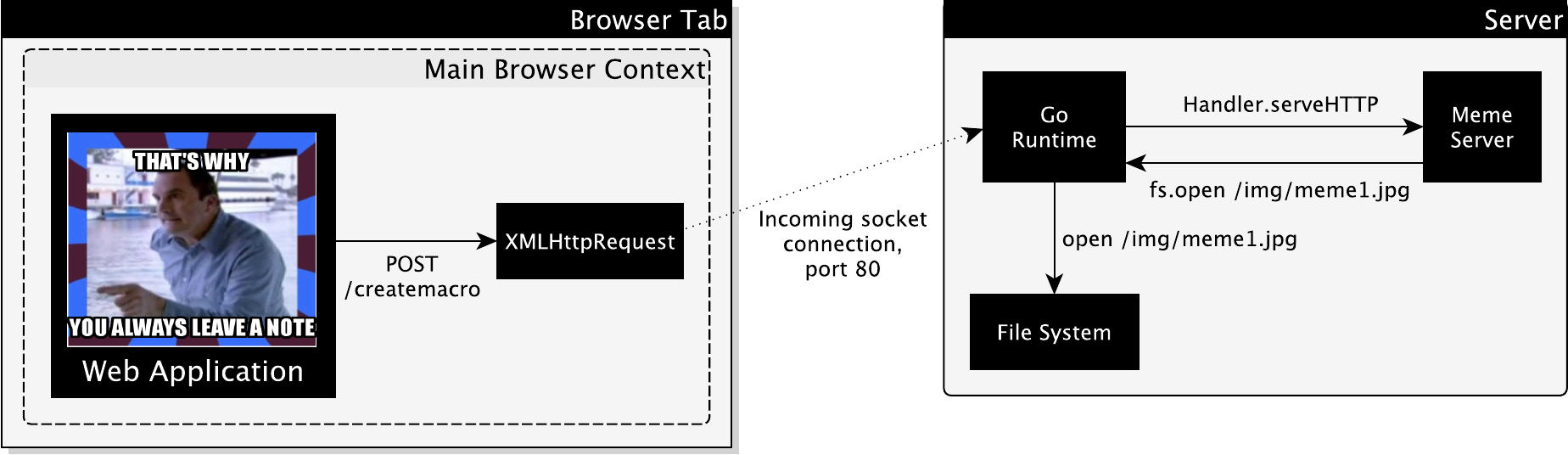}
    \label{fig:memegen-no-browsix}
  }

  \centering\subfloat[Meme creator running with \Browsix{}]{
    \includegraphics[width=.95\textwidth]{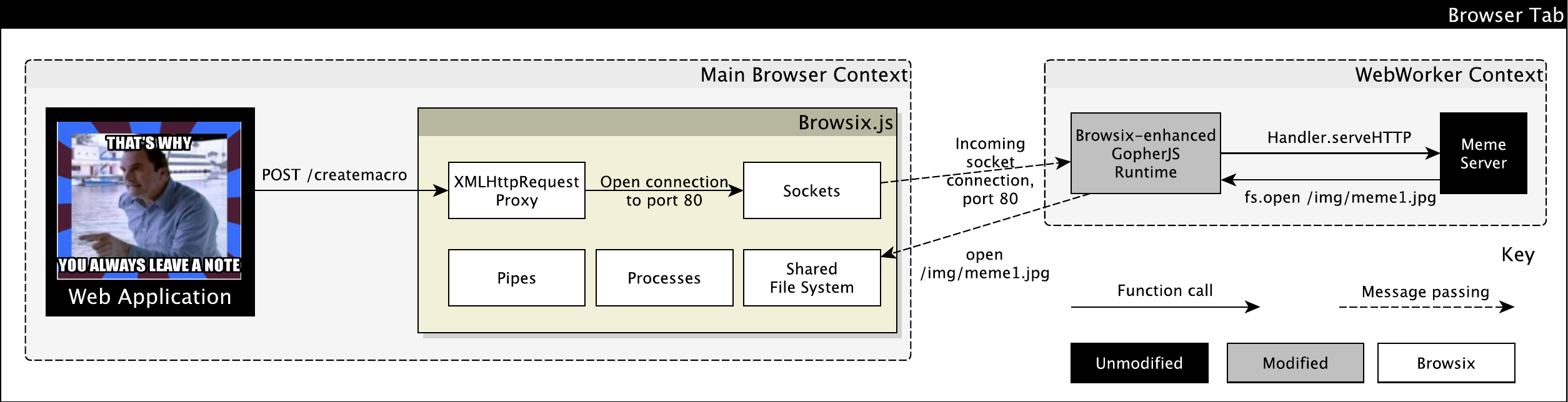}
    \label{fig:memegen-browsix}
  }
  \caption{System diagram of the meme generator application with and without \Browsix{}, demonstrating how the client and server interact with one another. With \Browsix{}, the server runs in the browser without code modifications.}
\end{figure*}

\subsubsection{Meme Generator:}
\label{subsubsec:image-macros}

Our meme generator lets users create \textit{memes} consisting of
images with (nominally) humorous overlaid
text. Figure~\ref{fig:imagemacrogen} contains a screenshot.  Existing
services, such as \url{MemeGenerator.net}, perform meme generation
server-side.  Moving meme creation into the browser would reduce
server load and reduce latency when the network is overloaded or
unreliable, but doing so would normally present a significant
engineering challenge.  The meme generation server uses sockets to
communicate with the browser over HTTP and reads meme templates from
the file system. Before \Browsix{}, the client and server code would
need to be re-architected and rewritten to run together in the
browser.

To demonstrate \Browsix{}'s ability to quickly port code from the
server to the web, we implement our meme creator as a traditional
client/server web application; Figure~\ref{fig:memegen-no-browsix}
contains a system diagram.  The client is implemented in HTML5 and
JavaScript, and the server is written in Go.  The server reads base
images and font files from the filesystem, and uses off-the-shelf
third-party Go libraries for image manipulation and font rendering to
produce memes~\cite{fogleman:2016gg}.  The server also uses Go's
built-in \texttt{http} module to run its web server. Note that this
server is stateless, following best
practices~\cite{mauro:2015netflix}; porting a stateful server would
naturally require more care.

To port the server to \Browsix{}, we follow the process outlined in
Section~\ref{sec:overview}.  First, we compile the Go server to a
single JavaScript file using GopherJS, a Go to JavaScript
compiler~\cite{musiol:2016gopherjs}.  Then, we stage the font and images for
the BrowserFS filesystem.  Finally, we augment the client application
to load the \Browsix JavaScript module, initialize a kernel instance,
and start the meme-server.

Next, we augment the web application to dynamically route meme
generation requests to a server running in \Browsix or to the cloud.
We add a function to the application that implements a simple policy:
if the network is inaccessible, or the browser is running on a desktop
(which is a proxy for a powerful device), the application routes meme
generation requests to the server running in \Browsix{}.  Otherwise,
it sends the requests to the remote server.  In both cases, the web
application uses an \texttt{XMLHttpRequest}-like interface to make the
request, requiring little change to the existing code.

Figure~\ref{fig:memegen-browsix} displays a system diagram of the
modified meme generator.  With this modification, meme generation
works even when offline. The code required to implement this policy
and dynamic behavior amounted to less than 30 lines of JavaScript.


\subsubsection{The \Browsix{} Terminal:}
\label{sec:shell}

To make it easy for developers to interact with and test programs in
\Browsix{}, we implement an in-browser Unix terminal that exposes a
POSIX shell. The terminal uses the Debian Almquist shell
(dash), the default shell of Debian and Ubuntu. We compile \texttt{dash} to JavaScript using
\Browsix-enhanced Emscripten, and run it in a \Browsix{} process.

Since the \Browsix{} terminal uses a standard Linux shell, developers
can use it to run arbitrary shell commands in \Browsix{}.
Developers can pipe programs together (e.g. \texttt{cat file.txt | grep apple > apples.txt}),
execute programs in a subshell in the background with \texttt{\&},
run shell scripts, and change environment variables.
Developers can also run Go, C/C++, and Node.js programs.

The terminal includes a variety of Unix utilities on the shell's PATH that we wrote for Node.js:
\texttt{cat}, \texttt{cp}, \texttt{curl}, \texttt{echo},
\texttt{exec}, \texttt{grep}, \texttt{head}, \texttt{ls},
\texttt{mkdir}, \texttt{rm}, \texttt{rmdir}, \texttt{sh},
\texttt{sha1sum}, \texttt{sort}, \texttt{stat}, \texttt{tail},
\texttt{tee}, \texttt{touch}, \texttt{wc}, and \texttt{xargs}.
These programs run equivalently under Node and \Browsix{} without any modifications,
and were used heavily during development to debug \Browsix{} functionality.

\paragraph{Summary:}
\Browsix{} makes it trivial to execute applications designed to run in
a Unix environment within the browser, enabling the rapid development
of sophisticated web applications.  These applications can incorporate
server code into the browser and harness the functionality of existing
applications.

\subsection{Performance}

We evaluate the performance overhead of \Browsix{} on our case
studies. All experiments were performed on a Thinkpad X1 Carbon with
an Intel i7-5600U CPU and 8 GB of RAM, running Linux 4.7.0.

\paragraph{\LaTeX{} Editor:}
Running \texttt{pdflatex} under \Browsix imposes an order of magnitude
slowdown.  A native build of \texttt{pdflatex} under Linux takes
around 100 milliseconds on a single page document with a bibliography.
When using synchronous calls (as supported by Google Chrome), the same
document builds in \Browsix in just under 3 seconds. While in relative
terms this is a significant slowdown, this time is fast enough to be
acceptable. Using asynchronous system calls and the Emterpreter, which is
only necessary to enable broader compatibility with today's browsers,
increases runtime to around 12 seconds.

\paragraph{Meme Generator:}
The meme generator performs two types of HTTP requests to the server:
requests for a list of available background
images, and requests to generate a meme.
We benchmark the performance of the meme generator server running
natively and running in Browsix in both Google Chrome and Firefox.

On average, a request for a list of background images takes 1.7 milliseconds natively,
9 ms in Google Chrome, and 6 ms in Firefox.
While requests to a server running natively on the same machine as the client are faster than
those served by \Browsix, \Browsix is
faster once a network connection and roundtrip latencies are factored
in.  When comparing an instance of the meme-server running on an EC2
instance, the in-\Browsix{} request completed three times as fast as
the request to the remote machine.  Times reported are the mean of 100
runs following a 20-run warmup.

The performance of meme generation is degraded by limitations in
current browsers.  The in-\Browsix HTTP request takes approximately
two seconds to generate a meme in the browser, versus 200 ms when
running server-side. This inefficiency is primarily due to missing
64-bit integer primitives when numerical code is compiled to
JavaScript with GopherJS, rather than overhead introduced by \Browsix;
we expect this to improve with both when future browsers support
native access to 64-bit integers, and with independent improvements to
the GopherJS compiler.

\paragraph{\Browsix{} Terminal and Utilities:}
Unix utilities provide a mechanism to compare the performance of
real-world programs under Linux and \Browsix.
Figure~\ref{fig:utilspeed} shows the results of running the same
JavaScript utility under \Browsix{} and on Linux under Node.js, and
compares this to the execution time of the corresponding GNU Coreutils
utility (written in C, running on Linux). Most of the overhead can be
attributed to JavaScript (the basis of Node.js and \Browsix{});
subsequently running in the \Browsix{} environment imposes roughly a
$3\times$ overhead over Node.js on Linux. Nonetheless, this
performance (completion in under 200 milliseconds) is low enough that
it should be generally acceptable to users.

\begin{figure}[!t]
\centering
\begin{tabular}{lrrr}
\textbf{Command}     & \textbf{Native} & \textbf{Node.js} & \textbf{\Browsix{}} \\
\hline
\small \texttt{sha1sum} & 0.002s & 0.067s & 0.189s \\
\small \texttt{ls}           & 0.001s & 0.044s & 0.108s \\
\end{tabular}
\caption{Execution time of utilities under \browsix{}, compared to the
  same utilities run under Node.js, and the native GNU/Linux
  utilities. \texttt{sha1sum} is run on \texttt{usr/bin/node}, and
  \texttt{ls} is run on \texttt{/usr/bin}. Running in JavaScript (with
  Node.js and \Browsix{}) imposes most overhead; running in the
  \Browsix{} environment adds roughly another $3\times$ overhead.}
\label{fig:utilspeed}
\end{figure}


\paragraph{Summary:} While \Browsix{}'s performance is limited by the performance of
underlying browser primitives (notably, the lack of native 64-bit
longs), it provides acceptable performance for a range of
applications.

\section{Discussion}
\label{sec:discussion}

\begin{table*}[ht!]
  \centering
    \begin{tabular}{>{\small}l>{\small}l>{\small}c>{\small}c>{\small}c>{\small}c>{\small}c>{\small}c>{\small}c}
    & & \rothead{\textbf{File system}} & \rothead{\textbf{Socket clients}} & \rothead{\textbf{Socket servers}} & \rothead{\textbf{Processes}} & \rothead{\textbf{Pipes}} & \rothead{\textbf{Signals}} \\
    \hline
    \textsc{Environments} & \Browsix{} & \cmark & \cmark & \cmark & \cmark & \cmark & \cmark \\
    & \doppio{}~\cite{doppio:pldi14}  & $\dagger$ & $\dagger$ & & & & \\
    & WebAssembly & & & & & & \\
    \hline
    \textsc{Language Runtimes} & Emscripten (C/C++) & $\dagger$ & $\dagger$ & $\dagger$ & & & & \\
    & GopherJS (Go) & & & & & & & \\
    & \Browsix{} + Emscripten & \cmark & \cmark & \cmark & \cmark & \cmark & \cmark \\
    & \Browsix{} + GopherJS & \cmark & \cmark & \cmark & \cmark & \cmark & \cmark \\
    \hline
  \end{tabular}
  \caption{ Feature comparison of JavaScript execution environments and language runtimes for programs compiled to JavaScript. $\dagger$ indicates that the feature is only accessible by a single running process. \Browsix{} provides multi-process support for all of its features.}

  \label{sec1:features}
\end{table*}

The process of implementing \Browsix{} has highlighted opportunities
for improvement in the implementation and specification of browser
APIs, especially Web Workers.  We outline a number of optimizations
and natural extensions that are generally useful, and would extend
\Browsix{}'s reach.

\paragraph{Worker Priority Control:}

The parent of a Web Worker has no way to lower the priority
of a created worker.  As workers are implemented on top of OS
threads, this concept maps cleanly onto OS-level
priorities/niceness. Providing this facility would let web
applications prevent a low-priority CPU-intensive worker from
interfering with the main browser thread.

%

\paragraph{\texttt{postMessage()} Backpressure:}
Traditional operating systems attempt to prevent individual processes
from affecting system stability in a number of ways.  One of these is
providing backpressure, wherein the process attempting to write to a
pipe or socket is suspended (the system call blocks) until the other
end of the pipe reads the data, or it can fit into a fixed size
buffer.  This approach prevents unbounded resource allocation in the
kernel.  In the browser, the \texttt{postMessage()} function can be
called from a JavaScript context an unbounded number of times and can eventually
cause the browser to run out of allocatable memory.

\paragraph{Message Passing Performance:}
Message passing is three orders of magnitude
slower than traditional system calls in the two browsers we evaluate,
Chrome and Firefox. A more efficient
message passing implementation would improve the performance of \Browsix{}'s
system calls and other inter-process communication.

\paragraph{Memory Mapping:}
\label{sec:system:limitations}
\Browsix{} is currently unable to support C/C++ applications like
PostgreSQL that use \texttt{mmap}.  Emscripten uses a single typed
array to represent the unmanaged C/C++ heap. While recent browser
interfaces make it possible to share this typed array among \Browsix{}
processes~\cite{tc39:2016sab}, browsers cannot yet map regions of the
typed array into another typed array, which would be necessary to fully
emulate \texttt{mmap}. Features on the WebAssembly roadmap, which aim
for implementation across browsers, would enable \Browsix{} to support
additional features like shared \texttt{mmap} and
\texttt{shm}\footnote{\url{https://github.com/WebAssembly/design/blob/master/FutureFeatures.md}}.

\paragraph{Stack Management:}
\label{par:stack-management}
C provides the ability to save and change the current thread's context
with the \texttt{setcontext} and \texttt{getcontext} functions.  While
rarely useful or advisable for applications, it enables
specialized low-level libraries to save and restore the C stack.  A
similar JavaScript primitive coupled with the use of
SharedArrayBuffers would let \Browsix{} support \texttt{fork} in Emscripten
applications as a synchronous system call.

\section{Related Work}
\label{sec:related}

\paragraph{In-Browser Execution Environments:}

\Browsix{} significantly extends past efforts to bring traditional
APIs and general-purpose languages to the browser;
Table~\ref{sec1:features} provides a comparison. Doppio's focus is
providing single-process POSIX
abstractions~\cite{doppio:pldi14}. \Browsix{} builds on and extends
its filesystem component, BrowserFS, to support multiple processes.
Emscripten compiles LLVM bytecode
to JavaScript, enabling the compilation of C and C++ to
JavaScript~\cite{emscripten}; as Section~\ref{sec:integrations}
describes, \Browsix{} augments its runtime system so that unmodified C
and C++ programs compiled with Emscripten can take full advantage of
its facilities. \Browsix{} provides similar runtime support for Go
programs through GopherJS~\cite{musiol:2016gopherjs}.


\paragraph{Kernel Design and OS Interfaces:}

\Browsix{} resembles a Linux kernel task running on a
microkernel~\cite{hartig1997performance}, as it relies on an underlying
system for messaging, scheduling and
context switching.  Barrelfish, a many-core, heterogenous
OS~\cite{baumann2009multikernel} showed that asynchronous,
shared-nothing system calls could be practical.  Additionally,
\Browsix somewhat mirrors the per-core, shared nothing structure of a
multikernel, as individual \Browsix processes do not use inter-domain
communication for tasks like memory allocation and timers.




\section{Conclusion}
\label{sec:conclusion}

This paper introduces \Browsix{}, a framework that brings the essence
of Unix to the browser.
\Browsix{} makes it almost trivial to build
complex web applications from components written in a variety of
languages without modifying any code, and promises to significantly
reduce the effort required to build highly sophisticated web
applications.  \Browsix{} is open source, and is freely available at
\url{https://github.com/plasma-umass/Browsix}.

\newpage

\bibliography{emery,main}

\begin{thebibliography}{10}

\bibitem{baumann2009multikernel}
A.~Baumann, P.~Barham, P.-E. Dagand, T.~Harris, R.~Isaacs, S.~Peter, T.~Roscoe,
  A.~Sch{\"u}pbach, and A.~Singhania.
\newblock The multikernel: a new {OS} architecture for scalable multicore
  systems.
\newblock In {\em Proceedings of the ACM SIGOPS 22nd symposium on Operating
  systems principles}, pages 29--44. ACM, 2009.

\bibitem{scala-js}
S.~Doeraene.
\newblock Scala.{j}s: {T}ype-{D}irected {I}nteroperability with {D}ynamically
  {T}yped {L}anguages.
\newblock Technical report, École polytechnique fédérale de Lausanne, 2013.

\bibitem{fogleman:2016gg}
M.~Fogleman.
\newblock {\em fogleman/gg: Go Graphics - 2D rendering in Go with a simple
  API}, 2016.
\newblock \url{https://github.com/fogleman/gg}.

\bibitem{tc39:2016sab}
L.~T. Hansen and J.~Fairbank.
\newblock {\em {ECMAScript} Shared Memory and Atomics}, 2016.
\newblock \url{https://tc39.github.io/ecmascript_sharedmem/shmem.html}.

\bibitem{hartig1997performance}
H.~H\"{a}rtig, M.~Hohmuth, J.~Liedtke, J.~Wolter, and S.~Sch\"{o}nberg.
\newblock The performance of {$\mu$}-kernel-based systems.
\newblock In {\em Proceedings of the Sixteenth ACM Symposium on Operating
  Systems Principles}, SOSP '97, pages 66--77, New York, NY, USA, 1997. ACM.

\bibitem{clojurescript}
R.~Hickey.
\newblock Clojurescript.
\newblock \url{http://clojurescript.org/about/rationale}, 2016.

\bibitem{mauro:2015netflix}
T.~Mauro.
\newblock {\em Adopting Microservices at Netflix: Lessons for Architectural
  Design}, 2015.
\newblock
  \url{https://www.nginx.com/blog/microservices-at-netflix-architectural-best-practices/}.

\bibitem{musiol:2016gopherjs}
R.~Musiol.
\newblock {\em gopherjs/gopherjs: A compiler from Go to JavaScript for running
  Go code in a browser}, 2016.
\newblock \url{https://github.com/gopherjs/gopherjs}.

\bibitem{doppio:pldi14}
J.~Vilk and E.~D. Berger.
\newblock {\sc Doppio}: Breaking the browser language barrier.
\newblock In {\em Proceedings of the 2014 ACM SIGPLAN Conference on Programming
  Language Design and Implementation (PLDI 2014)}, pages 508--518. ACM, 2014.

\bibitem{js-of-ocaml}
J.~Vouillon and V.~Balat.
\newblock {From bytecode to JavaScript: the Js\_of\_ocaml compiler}.
\newblock {\em Software: Practice and Experience}, 44(8):951--972, 2014.

\bibitem{whalesong}
D.~Yoo and S.~Krishnamurthi.
\newblock Whalesong: running racket in the browser.
\newblock In {\em DLS'13, Proceedings of the 9th Symposium on Dynamic
  Languages, part of {SPLASH} 2013, Indianapolis, IN, USA, October 26-31,
  2013}, pages 97--108, 2013.

\bibitem{emscripten}
A.~Zakai.
\newblock Emscripten: an {LLVM}-to-{JavaScript} compiler.
\newblock In {\em OOPSLA Companion}, pages 301--312, 2011.

\end{thebibliography}
\bibliographystyle{abbrv}


\end{document}